\documentclass[fleqn,10pt]{wlscirep}

\usepackage{verbatim}
\usepackage{dsfont}
\usepackage{cleveref}

\newcommand{\bra}[1]{\left\langle #1\right|}
\newcommand{\ket}[1]{\left| #1\right\rangle}

\newcommand{\opav}[3]{\langle #1 | #2 | #3 \rangle}

\newcommand{\beq}{\begin{equation}}
\newcommand{\eeq}{\end{equation}}

\newcommand{\tr}{\text{Tr}}
\newcommand{\tot}[1]{#1_{\text{tot}}}

\begin{document}

\title{The quantum Zeno and anti-Zeno effects with strong system-environment coupling}

\author[1,*]{Adam Zaman Chaudhry}
\affil[1]{School of Science \& Engineering, Lahore University of Management Sciences (LUMS), Opposite Sector U, D.H.A, Lahore 54792, Pakistan}

\affil[*]{adam.zaman@lums.edu.pk}

\begin{abstract}

The fact that repeated projective measurements can slow down (the Zeno effect) or speed up (the anti-Zeno effect) quantum evolution is well-known. However, to date, studies of these effects focus on quantum systems that are weakly interacting with their environment. In this paper, we investigate what happens to a quantum system under the action of repeated measurements if the quantum system is strongly interacting with its environment. We consider as the quantum system a single two-level system coupled strongly to a collection of harmonic oscillators. A so-called polaron transformation is then used to make the problem in the strong system-environment coupling regime tractable. We find that the strong coupling case exhibits quantitative and qualitative differences as compared with the weak coupling case. In particular, the effective decay rate does not depend linearly on the spectral density of the environment. This then means that, in the strong coupling regime that we investigate, increasing the system-environment coupling strength can actually decrease the effective decay rate. We also consider a collection of two-level atoms coupled strongly with a common environment. In this case, we find that there are further differences between the weak and strong coupling cases since the two-level atoms can now indirectly interact with one another due to the common environment.

\end{abstract}

\maketitle

By repeatedly measuring a quantum system very frequently, the evolution of the quantum system can be slowed down, an effect that has been dubbed as the Quantum Zeno effect (QZE)\cite{Sudarshan1977,FacchiPhysLettA2000,FacchiPRL2002, FacchiJPA2008, WangPRA2008, ManiscalcoPRL2008, FacchiJPA2010, MilitelloPRA2011, RaimondPRA2012, SmerziPRL2012, WangPRL2013, McCuskerPRL2013, StannigelPRL2014, ZhuPRL2014, SchafferNatCommun2014,SignolesNaturePhysics2014,DebierrePRA2015,AlexanderPRA2015,QiuSciRep2015,Slichterarxiv2015}. On the other hand, if the quantum system is measured repeatedly not very rapidly, the measurements can actually speed up the temporal evolution. This effect, the opposite of the QZE, is known as the Quantum anti-Zeno effect (QAZE)\cite{KurizkiNature2000, KoshinoPhysRep2005,BennettPRB2010,BaronePRL2004,YamamotoPRA2010,RaizenPRL2001}. Both the QZE and the QAZE have attracted tremendous theoretical and experimental interest due to their great importance for emerging quantum technologies as well as their fundamental theoretical interest. However, it is worth noting that the emphasis in studies performed on the QZE and the QAZE to date has been on the population decay of quantum systems. In these studies, the quantum system is prepared in an excited state, and then the system is repeatedly checked to see if the system is still in the excited state or not\cite{KurizkiNature2000, KoshinoPhysRep2005,BennettPRB2010,BaronePRL2004,YamamotoPRA2010,RaizenPRL2001,ManiscalcoPRL2006,SegalPRA2007,ZhengPRL2008,
AiPRA2010,ThilagamJMP2010,ThilagamJCP2013}. It is well-known then that the decay rate of the quantum system depends on the overlap of the spectral density of the environment and a measurement-induced level width\cite{KurizkiNature2000}. Depending on this overlap, decreasing the measurement interval can lead to a decrease (the QZE) or an increase (the QAZE) of the decay rate. 

While studies of the QZE and the QAZE performed to date by and large focus on the population decay model where only decay takes place, we also know from the study of open quantum systems that, in general, quantum systems interacting with their environment also undergo dephasing. To this end, the QZE and the QAZE were studied for the exactly solvable pure dephasing model in Ref.~{\renewcommand{\citemid}{}\cite[]{ChaudhryPRA2014zeno}}, where it was shown that the QZE and the QAZE are significantly different for the pure dephasing case as compared with the population decay case. This study was then extended to arbitrary system-environment models in Ref.~{\renewcommand{\citemid}{}\cite[]{Chaudhryscirep2016}} where a general framework for calculating the effective decay rate of the system for an arbitrary system-environment model was presented. It was found that the effective decay rate can be written as an overlap integral of the spectral density of the environment and an effective `filter function' that depends on the system-environment model at hand, the measurement interval, and the measurement being repeatedly performed. This general formalism was then used to study the QZE and the QAZE when both dephasing and population decay are present. For example, repeated measurements for the paradigmatic spin-boson model \cite{LeggettRMP1987} were considered and it was shown that the presence of both population decay and dephasing make the results differ considerably both quantitatively and qualitatively as compared to the pure population decay case. 

It should be pointed out, however, that the results presented in Ref.~{\renewcommand{\citemid}{}\cite[]{Chaudhryscirep2016}} were derived under the assumption that the system-environment coupling is weak. This is consistent with studies performed for the population decay models, where the effective decay rate can be derived using time-dependent perturbation theory \cite{KoshinoPhysRep2005}. On the other hand, the behavior of a quantum system, subjected to repeated measurements, that is interacting strongly with its environment is not well understood. For instance, one could ask whether or not the effective decay rate is still an overlap integral of the spectral density function and a `filter' function. This paper intends to answer precisely such questions by looking at what happens to the spin-boson model under the action of repeated measurements if the central two-level system is interacting strongly with a surrounding environment of harmonic oscillators. Since the system-environment coupling is strong, the system-environment interaction cannot be treated perturbatively, and thus the treatment given in Ref.~{\renewcommand{\citemid}{}\cite[]{Chaudhryscirep2016}} is no longer applicable. Our strategy then is to perform a unitary transformation, known as the polaron transformation, on the system-environment Hamiltonian \cite{SilbeyJCP1984,VorrathPRL2005,ChinPRL2011,LeeJCP2012,LeePRE2012,GuzikJPCL2015}.  One then finds that the system and the environment can end up interacting weakly in this new `polaron' frame. Perturbation theory can then be applied and the effect of repeated measurements is analyzed. We find that the analysis of the QZE and QAZE are in general very different compared to the population decay case. For example, it is clear that for the usual population decay case, increasing the system-environment strength increases the effective decay rate. However, for the strong system-environment regime that we investigate, we find that increasing the system-environment coupling regime can actually decrease the effective decay rate. We also study the QZE and the QAZE for more than one two-level system interacting with a common environment. For the weak coupling regime, the effective decay rate is directly proportional to the number of two-level systems coupled to the common environment \cite{Chaudhryscirep2016}. On the other hand, for the strong system-environment coupling regime, we find that the effective decay rate for more than one two-level system is very different compared to the single two-level system case. The indirect interaction between the two-level systems due to their interaction with a common environment now plays a very important role, and the effective decay rate is no longer simply proportional to the number of two-level systems coupled to the common environment.  

\section*{Results}
\subsection*{Spin-boson model with strong system-environment coupling}
We start with the paradigmatic spin-boson model Hamiltonian \cite{LeggettRMP1987,Weissbook,BPbook} which we write as (we set $\hbar = 1$ throughout)
\begin{equation}
H_L = \frac{\varepsilon}{2} \sigma_z + \frac{\Delta}{2} \sigma_x + \sum_k \omega_k b_k^\dagger b_k + \sigma_z \sum_k (g_k^* b_k + g_k b_k^\dagger), 
\end{equation}
where the system Hamiltonian is $H_{S,L} = \frac{\varepsilon}{2} \sigma_z + \frac{\Delta}{2} \sigma_x $, the environment Hamiltonian is $H_B = \sum_k \omega_k b_k^\dagger b_k$, and the system-environment coupling is $V_L = \sigma_z \sum_k (g_k^* b_k + g_k b_k^\dagger)$. $\varepsilon$ is the energy level difference of the two-level system, $\Delta$ is the tunneling amplitude, $\omega_k$ are the frequencies of the harmonic oscillators, $b_k$ and $b_k^\dagger$ are the annihilation and creation operators for the harmonic oscillators, and $\sigma_{x}$ and $\sigma_z$ are the standard Pauli operators. The `$L$' denotes the `lab' frame. If the system-environment coupling is strong, we cannot treat the system-environment coupling perturbatively. Furthermore, the system-environment correlation effects are significant as well in general. To motivate our basic approach in this strong coupling regime, we note that if the system tunneling amplitude is negligible and the initial system state is an eigenstate of $\sigma_z$, then, even though the system and the environment are strongly interacting, the evolution of the system state is negligible. This then means that we should look to unitarily transform $H_L$ such that the effective system-environment coupling contains the tunneling amplitude $\Delta$. This unitary transformation is provided by the `polaron' transformation, whereby the system-environment Hamiltonian in this new `polaron' frame becomes $H = e^{\chi \sigma_z/2} H_L e^{-\chi \sigma_z/2}$, where $\chi = \sum_k \left[ \frac{2 g_k}{\omega_k} b_k^\dagger - \frac{2g_k^*}{\omega_k}b_k \right]$ \cite{SilbeyJCP1984,VorrathPRL2005,ChinPRL2011,LeeJCP2012,LeePRE2012,GuzikJPCL2015}. The system-environment Hamiltonian in the polaron frame is then $ H = H_S + H_B + V$, where $H_S = \frac{\varepsilon}{2} \sigma_z$, $H_B = \sum_k \omega_k b_k^\dagger b_k$, and $V = \frac{\Delta}{2} \left[ \sigma_+ X + \sigma_- X^\dagger \right]$, with $X = e^\chi$ (see Methods for details). Now, if the tunneling amplitude is small, we can use time-dependent perturbation theory, treating $V$ as the perturbation. This is the key idea to deal with the strong system-environment coupling regime. Although the system and the environment are strongly interacting, in the polaron frame, they are effectively interacting weakly. Let us now use this fact in order to calculate the survival probability, and thereby the effective decay rate. For concreteness, we assume that the initial state prepared is $\ket{\uparrow}$, where $\sigma_z \ket{\uparrow} = \ket{\uparrow}$. In other words, we consider the same initial state as that considered in the analysis of the usual population decay model \cite{KurizkiNature2000,KoshinoPhysRep2005}. At time $t = 0$, we prepare the system state $\ket{\uparrow}$, and we subsequently perform measurements with time interval $\tau$ to check if the system state is still $\ket{\uparrow}$ or not. The survival probability after time interval $\tau$ is then $s(\tau) = \tr_{S,B} [\ket{\uparrow}\bra{\uparrow} \rho_L(\tau)]$,
where $\rho_L(\tau)$ is the combined density matrix of the system and the environment at time $\tau$ just before the projective measurement. Then, 
$$s(\tau) = \text{Tr}_{S,B} [\ket{\uparrow}\bra{\uparrow} e^{-iH_L \tau} \rho_L(0) e^{iH_L \tau}]. $$
It is important to note that the initial state that we have prepared cannot simply be taken as the usual product state $\ket{\uparrow}\bra{\uparrow} \otimes e^{-\beta H_B}/Z_B$, with $Z_B = \tr_B [e^{-\beta H_B}]$ since the system and the environment are strongly interacting and consquently there will be significant initial system-environment correlations \cite{ChaudhryPRA2013a,ChaudhryPRA2013b} . Rather, the initial state that we should consider is $\rho_L(0) = P_\uparrow e^{-\beta H_L} P_\uparrow /Z$, where $P_\uparrow = \ket{\uparrow}\bra{\uparrow}$, and $Z = \tr_{S,B} [P_\uparrow e^{-\beta H_L}]$. Keeping this in mind, we use the polaron transformation to cast the expression for the survival probability $s(\tau)$ after the measurement at time $\tau$ in terms of quantities in the polaron frame. Doing so leads us to  
$$ s(\tau) = \tr_{S,B} [\ket{\uparrow}\bra{\uparrow} e^{-iH\tau} P_\uparrow \frac{e^{-\beta H}}{Z} P_\uparrow e^{iH\tau}],$$
with the Hamiltonian $H$ now in the polaron frame. Now, for small $\Delta$, to a first approximation, the initial state in the polaron frame can be written as $P_\uparrow \otimes e^{-\beta H_B}/Z_B$. This is a similar approximation as the usual assumption that the initial system-environment state is $\rho_S(0) \otimes \rho_B$ since, in the polaron frame, the system and the environment are weakly interacting. We thus get
$$ s(\tau) = \tr_{S,B} [\ket{\uparrow}\bra{\uparrow} e^{-iH\tau} (\ket{\uparrow} \bra{\uparrow} \otimes e^{-\beta H_B}/Z_B) e^{iH\tau}] = \tr_S[\ket{\uparrow}\bra{\uparrow} \rho_S(\tau)],$$
where $\rho_S(\tau) = \tr_B [e^{-iH\tau} (\ket{\uparrow} \bra{\uparrow} \otimes e^{-\beta H_B}/Z_B) e^{iH\tau}]$. Our objective then is to find $\rho_S(\tau)$, given the initial system-environment state $\rho(0) = \ket{\uparrow} \bra{\uparrow} \otimes e^{-\beta H_B}/Z_B$. We find that (see the Methods section)
\begin{align}
			\rho_S(\tau) = &U_S(\tau) \biggl( \rho_S(0) + i \sum_\mu \int_0^\tau dt_1 [\rho_S(0),\widetilde{F}_\mu(t_1)] \langle \widetilde{B}_\mu(t_1) \rangle_B + \notag \\
			&\sum_{\mu \nu}\int_0^\tau dt_1 \int_0^{t_1} dt_2 \bigl\lbrace C_{\mu \nu}(t_1, t_2) 
			[\widetilde{F}_\nu(t_2)\rho_S(0),\widetilde{F}_\mu(t_1)] + \text{h.c.}\bigr\rbrace \biggr) U_S^\dagger (\tau).
			\label{densitymatrixattau}
			\end{align}
Here $U_S(\tau) = e^{-iH_S\tau}$, $F_1 = \frac{\Delta}{2}\sigma_+$, $B_1 = X$, $F_2 = \frac{\Delta}{2}\sigma_-$, $B_2 = X^\dagger$, $\widetilde{F}_\mu(t) = U_S^\dagger(t) F_\mu U_S(t)$, $\widetilde{B}_\mu (t) = U_B^\dagger (t) B_\mu U_B (t)$ with $U_B(t) = e^{-iH_B t}$, $\langle \hdots \rangle_B = \tr_B[\rho_B (\hdots)]$ where $\tr_B$ denotes taking trace over the environment, the environment correlation functions are defined as $C_{\mu \nu}(t_1,t_2) = \langle \widetilde{B}_\mu(t_1)\widetilde{B}_\nu(t_2)\rangle_B$, and h.c. denotes the hermitian conjugate. Now, since the system-environment coupling in the polaron frame is weak, we can neglect the build up of correlations between the system and the environment. Thus, we can write the survival probability after time $t = N\tau$, where $N$ is the number of measurements performed after time $t = 0$, as $S(t = N\tau) = [s(\tau)]^N \equiv e^{-\Gamma(\tau)N\tau}$, thereby defining the effective decay rate $\Gamma(\tau)$. It then follows that $\Gamma(\tau) = -\frac{1}{\tau} \ln s(\tau)$. Since we have the system density matrix in the polaron frame, we can work out the survival probability $s(\tau)$ and hence the effective decay rate $\Gamma(\tau)$. The result is that (see the Methods section for details)
			\begin{align}
			\Gamma(\tau) &= \frac{\Delta^2}{2\tau} \int_0^\tau dt \int_0^t dt'  e^{-\Phi_R(t')} \cos[\varepsilon t' - \Phi_I(t')],
			\label{decayrateoneTLS}
			\end{align}
			where 
			\begin{align}
			\Phi_R(t) = \int_0^\infty \, d\omega \,J(\omega) \frac{1 - \cos(\omega t)}{\omega^2} \coth\left(\frac{\beta \omega}{2}\right), \;
			\Phi_I(t) = \int_0^\infty \, d\omega \,J(\omega) \frac{\sin(\omega t)}{\omega^2},
			\label{PhiRandPhiI}
			\end{align}
			and the spectral density of the environment has been introduced as $\sum_k |g_k|^2 (\hdots) \rightarrow \int_0^\infty d\omega J(\omega)(\hdots)$. At this point, it is useful to compare this expression for the effective decay rate for the case of strong system-environment coupling with the case of the usual population decay model where the effective decay rate is $\Gamma(\tau) = \tau \int_0^\infty \, d\omega\, J(\omega) \text{sinc}^2 \left[\frac{(\varepsilon - \omega)\tau}{2}\right]$\cite{KurizkiNature2000,KoshinoPhysRep2005}. It should be clear that for the strong system-environment coupling case, the effective decay rate given by Eq.~\eqref{decayrateoneTLS} has a very different qualitative behavior. In particular, the effective decay rate can no longer be regarded as simply an overlap integral of the spectral density of the environment with a sinc-squared function. Rather, the effective decay rate now has a very prominent non-linear dependence on the spectral density, leading to very different behavior as compared with the population decay case. For example, as the system-environment coupling strength increases, $\Phi_R(t)$ increases, and thus we expect $\Gamma(\tau)$ to decrease. To make this claim concrete, let us model the spectral density  as $J(\omega) = G \omega^s \omega_c^{1 - s} e^{-\omega/\omega_c}$, where $G$ is a dimensionless parameter characterizing the system-environment coupling strength, $\omega_c$ is the cutoff frequency, and $s$ is the Ohmicity parameter \cite{BPbook}. For concreteness, we look at the Ohmic case $(s = 1)$. In this case, $ \Phi_R(t) = \frac{G}{2} \ln (1 + \omega_c^2 t^2)$, while $\Phi_I(t) = G \tan^{-1} (\omega_c t)$, leading to 
			\begin{align}
			\Gamma(\tau) &= \frac{\Delta^2}{2\tau} \int_0^\tau dt \int_0^t dt'  \frac{\cos[\varepsilon t' - G \tan^{-1}(\omega_c t')]}{(1 + \omega_c^2 t'^2)^{G/2}}. 
			\end{align}
			
\begin{figure}[t!]
\centering
{\includegraphics[scale = 0.5]{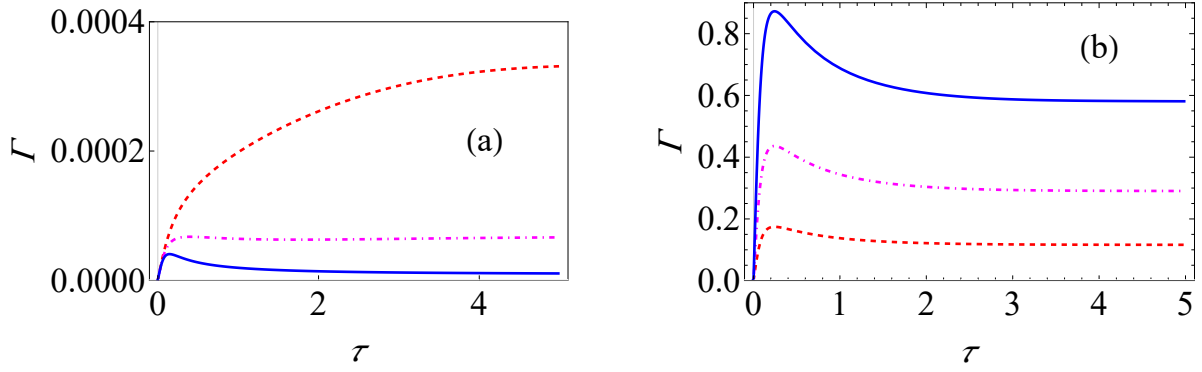}}
\caption{\textbf{Variation of the effective decay rate with change in system-environment coupling strength.} (a) Graph of $\Gamma$ (at zero temperature) for the strong-coupling regime as a function of $\tau$ with the system-environment coupling strength $G = 1$ (red, dashed curve), $G = 1.75$ (dot-dashed, magenta curve), and $G = 2.5$ (solid, blue curve). Here we have used an Ohmic environment ($s = 1$), with $\varepsilon = 1$, $\omega_c = 10$, and $\Delta = 0.05$. The initial state is $\ket{\uparrow}$. (b) Behaviour of $\Gamma$ (at zero temperature) for the usual weak system-environment coupling scenario leading to only population decay for $G = 0.02$ (dashed, red curve), $G = 0.05$ (dot-dashed, magenta curve), and $G = 0.1$ (solid, blue curve). Here we have used again an Ohmic environment, the initial state is still $\ket{\uparrow}$, $\varepsilon = 1$, and $\omega_c = 10$. Throughout, we use dimensionless units with $\hbar = 1$.}
\label{labelohmiccouplingstrengthvariation}
\end{figure}

\begin{figure}[t!]
\centering
{\includegraphics[scale = 0.5]{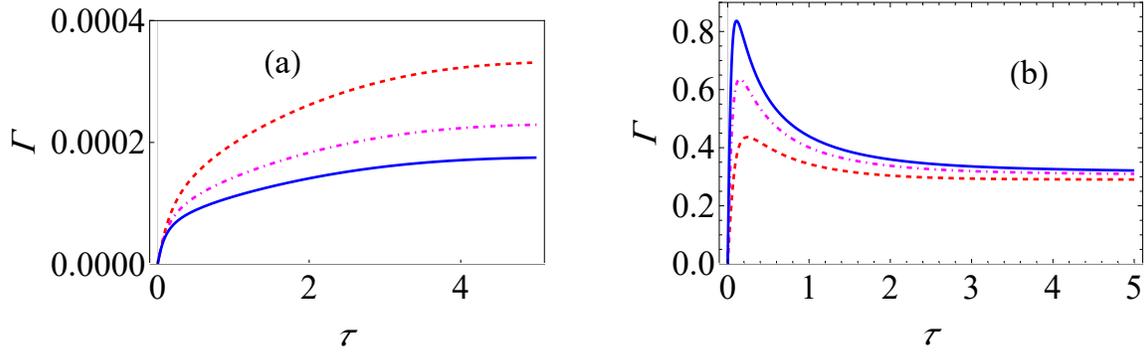}}
\caption{\textbf{Variation of the effective decay rate with change in the cutoff frequency.} (a) Graph of $\Gamma$ (at zero temperature) for the strong-coupling regime as a function of $\tau$ with $\omega_c = 10$ (red, dashed curve), $\omega_c = 15$ (dot-dashed, magenta curve), and $\omega_c = 20$ (solid, blue curve). Here we have used an Ohmic environment ($s = 1$), with $\varepsilon = 1$, $G = 1$, and $\Delta = 0.05$. The initial state is $\ket{\uparrow}$. (b) Behaviour of $\Gamma$ (at zero temperature) for the usual weak system-environment coupling scenario leading to only population decay for $\omega_c = 10$ (red, dashed curve), $\omega_c = 15$ (dot-dashed, magenta curve), and $\omega_c = 20$ (solid, blue curve). Here we have used again an Ohmic environment, the initial state is still $\ket{\uparrow}$, $\varepsilon = 1$, and $G = 0.05$.}
\label{labelohmiccouplingcutoffvariation}
\end{figure}

			The double integral can be worked out numerically. Results are shown in Fig.~\ref{labelohmiccouplingstrengthvariation}(a) for different system-environment coupling strengths $G$. For the strong system-environment regime that we are dealing with, it is clear that increasing the system-environment coupling strength $G$ actually decreases the effective decay rate. This is in contrast with what happens in the weak system-environment regime for the paradigmatic population decay model [see Fig.~\ref{labelohmiccouplingstrengthvariation}(b)]. Here it is clear that increasing the system-environment coupling strength increases the effective decay rate as expected. It should also be noted that the behaviour of $\Gamma(\tau)$ as a function of $\tau$ allows us to identify the Zeno and anti-Zeno regimes. One approach is to simply say that if $\Gamma(\tau)$ decreases when $\tau$ decreases, we are in the Zeno regime, while if $\Gamma(\tau)$ increases if $\tau$ decreases, then we are in the anti-Zeno regime \cite{KurizkiNature2000, SegalPRA2007, ThilagamJMP2010, ChaudhryPRA2014zeno}. From Fig.~\ref{labelohmiccouplingstrengthvariation}(b), it should also be noted that increasing the coupling strength does not change the qualitative behavior of the Zeno to anti-Zeno transition, but for the strong coupling regime [see Fig.~\ref{labelohmiccouplingstrengthvariation}(a)], while we only observe the Zeno effect for $G = 1$, both the Zeno and anti-Zeno effects are observed for $G = 2.5$. Similarly, as shown in Fig.~\ref{labelohmiccouplingcutoffvariation}(a), increasing the cutoff frequency for the strong coupling case decreases the effective decay rate, but the opposite behaviour is observed for the weak coupling case [see Fig.~\ref{labelohmiccouplingcutoffvariation}(b)].

\begin{figure}
\centering
{\includegraphics[scale = 0.5]{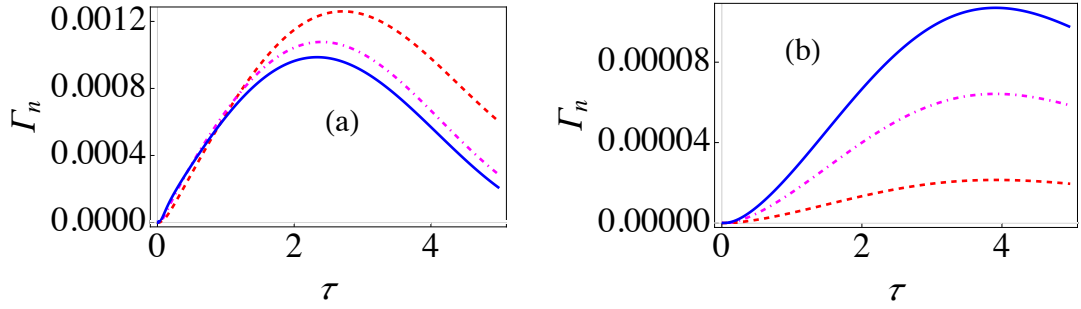}}
\caption{\textbf{Dependence of the modified decay rate $\Gamma_n(\tau)$ on the system-environment coupling strength.} (a) Graph of the effective decay rate $\Gamma_n(\tau)$ (at zero temperature) in the strong system-environment coupling regime as a function of $\tau$ with the system-environment coupling strength $G = 1$ (red, dashed curve), $G = 1.75$ (dot-dashed, magenta curve), and $G = 2.5$ (solid, blue curve). Here we have used an Ohmic environment ($s = 1$), with $\varepsilon = 1$, $\omega_c = 10$, and $\Delta = 0.05$. The initial state is $\ket{\uparrow}$. (b) Behaviour of $\Gamma$ (at zero temperature) for the usual weak system-environment coupling scenario with $\varepsilon = 1$ and $\Delta = 0.05$ with $G = 0.001$ (dashed, red curve), $G = 0.003$ (dot-dashed, magenta curve), and $G = 0.005$ (solid, blue curve). We have $s = 1$, the initial state is still $\ket{\uparrow}$ and $\omega_c = 10$.}
\label{labelspinhalfgraphswithcorrections}
\end{figure}

In our treatment until now, we have considered the change in the system state due to the tunneling term. This tunneling term, due to its presence in $H_{S,L} = \frac{\varepsilon}{2}\sigma_z + \frac{\Delta}{2}\sigma_x$, leads to the system state changing even if the system and the environment are not coupled to each other. Thus, an alternative way to quantify the effective decay rate would be to remove the evolution due to the system Hamiltonian (in the `lab' frame) $H_{S,L}$ before performing each measurement since what we are really interested in is the change in the system state due to the system-environment interaction. A similar approach has been followed in Refs.~{\renewcommand{\citemid}{}\cite[]{MatsuzakiPRB2010,ChaudhryPRA2014zeno,Chaudhryscirep2016}}. Therefore, we now derive an expression for the effective decay rate of the system state when, just before each measurement, we remove the system evolution due to $H_{S,L}$. The survival probability, after one measurement, is now (starting from the state $\ket{\uparrow}$)
\begin{equation}
s(\tau) = \tr_{S,B} [ (\ket{\uparrow}\bra{\uparrow}) e^{iH_{S,L}\tau} e^{-iH_L \tau} \rho_{\text{L}}(0) e^{iH_L \tau} e^{-iH_{S,L}\tau}]. 
\label{survivalprobabilitywithmod}
\end{equation}
Notice now the presence of $e^{iH_{S,L}\tau}$ and $e^{-iH_{S,L}\tau}$ which remove the evolution of the system due to the system Hamiltonian before performing the measurement. Once again transforming to the polaron frame, we obtain
$$ s(\tau) = 1 - \tr_{S,B} [ (\ket{\downarrow}\bra{\downarrow}) e^{iH_{S,P}\tau} e^{-iH \tau} (\ket{\uparrow}\bra{\uparrow} \otimes \rho_B) e^{iH \tau} e^{-iH_{S,P}\tau}], $$
where $H_{S,P} = \frac{\varepsilon}{2}\sigma_z + \frac{\Delta}{2}(\sigma_+ X + \sigma_- X^\dagger)$ and $\rho_B = e^{-\beta H_B}/Z_B$. Since we are assuming that the tunneling amplitude is small, the unitary operator $e^{-iH_{S,P}\tau}$ can be expanded as a perturbation series. At the same time, $e^{-iH\tau}$ can also expanded as a perturbation series. Keeping terms to second order in the tunneling amplitude (see the Methods section), we find that now the modified decay rate $\Gamma_n(\tau)$ is 
\begin{equation}
\Gamma_n(\tau) = \Gamma(\tau) + \Gamma_{\text{mod}}(\tau), 
\label{modifieddecayratesingleTLS}
\end{equation}
where the modification to the previous decay rate is 
$$ \Gamma_{\text{mod}}(\tau) = \frac{\Delta^2}{\tau} \left\lbrace \frac{1}{\varepsilon^2} \sin^2 \left(\frac{\varepsilon \tau}{2} \right) e^{-\Phi_R(0)} e^{-i\Phi_I(0)} - \frac{1}{\varepsilon} \sin\left( \frac{\varepsilon \tau}{2} \right) \int_0^\tau dt e^{-\Phi_R(t)} \cos \left[\varepsilon \left(t - \frac{\tau}{2}\right) - \Phi_I(t)\right]\right\rbrace. $$
Using these expressions, we have plotted the behavior of $\Gamma_n(\tau)$ for the strong system-environment coupling regime in Fig.~\ref{labelspinhalfgraphswithcorrections}(a). It should be clear that once again increasing the system-environment coupling strength generally decreases the effective decay rate $\Gamma_n(\tau)$. This is in sharp contrast with what happens in the weak coupling regime. For the weak coupling case, it is known that \cite{Chaudhryscirep2016}
\begin{equation}
\label{weakcouplingdecay}
\Gamma_n(\tau) = \int_0^\infty d\omega \, J(\omega) Q(\omega,\tau),
\end{equation}
where the filter function $Q(\omega,\tau)$ is 
\begin{equation}
			Q(\omega,\tau) = \frac{2}{\tau}\left\lbrace \coth \left(\frac{\beta \omega}{2}\right) D_1(\omega,\tau)  + D_2(\omega,\tau)\right\rbrace,
			\label{weakcouplingfilterfunction}
			\end{equation}
			with $D_1(\omega,\tau) = \int_0^\tau dt \int_0^t dt' \cos (\omega t') [a_x(t - t') a_x(t) + a_y(t - t')a_y(t)]$, and $D_2(\omega,\tau) = \int_0^\tau dt \int_0^t dt' \sin(\omega t') [-a_x(t - t')a_y(t) + a_x(t) a_y(t - t')]$.
			Here $a_x(t) = \frac{2\varepsilon \Delta}{\Omega^2} \sin^2 \left( \frac{\Omega t}{2} \right)$ and $a_y(t) = \frac{\Delta}{\Omega} \sin(\Omega t)$ with $\Omega^2 = \varepsilon^2 + \Delta^2$. Using these expressions, we can investigate how the decay rate varies as the measurement interval changes for different system-environment coupling strengths in the weak coupling regime. Typical results are illustrated in Fig.~\ref{labelspinhalfgraphswithcorrections}(b) from which it should be clear that increasing the coupling strength in the weak coupling regime increases the effective decay rate. Furthermore, changing the coupling strength has no effect on the measurement time interval at which the Zeno to anti-Zeno transition takes place for the weak coupling regime as the three curves in Fig.~\ref{labelspinhalfgraphswithcorrections}(b) achieve their maximum value for the same value of $\tau$. This is not the case for the strong coupling regime [see Fig.~\ref{labelspinhalfgraphswithcorrections}(a)].
			
			At this point, it is worth pausing to consider where the qualitative difference in the behavior of the effective decay rate in the weak and the strong coupling regime comes from. The effective decay rate is derived from the survival probability after one measurement $s(\tau)$. For both the weak and the strong coupling regimes, the survival probability after one measurement is given by Eq.~\eqref{survivalprobabilitywithmod}. For both cases, the Hamiltonian $H_{S,L}$ and $H_L$ are the same. The only difference is in the choice of the system-environment state $\rho_L(0)$. For the weak coupling case, this state is simply the product state $\ket{\uparrow}\bra{\uparrow}\otimes e^{-\beta H_B}/Z_B$. This is not the case for the strong coupling due to the significant system-environment correlations. Thus, we can say that the qualitative difference in the behavior of the effective decay rate is because of the presence of the system-environment correlations. It seems that these correlations can protect the quantum state of the system - as the coupling strength increases, these correlations become more and more significant, and at the same time, the effective decay rate goes down.
			
\subsection*{Large spin-boson model with strong system-environment coupling}

\begin{figure}
\centering
{\includegraphics[scale = 0.7]{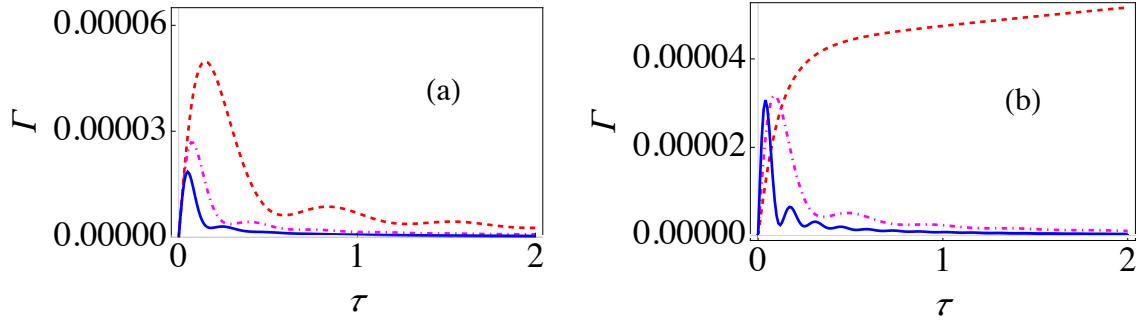}}
\caption{\textbf{Variation of the decay rate $\Gamma(\tau)$ as the coupling strength and the number of two-level systems is changed.} (a) Graph of $\Gamma$ (at zero temperature) for the strong-coupling regime with $j = 1$ as a function of $\tau$ with the system-environment coupling strength $G = 1$ (red, dashed curve), $G = 1.75$ (dot-dashed, magenta curve), and $G = 2.5$ (solid, blue curve). Here we have used an Ohmic environment ($s = 1$), with $\varepsilon = 1$, $\omega_c = 10$, and $\Delta = 0.05$. The initial state is $\ket{j}$. (b) Graph of $\Gamma$ (at zero temperature) for the strong-coupling regime with $G = 1.5$ as a function of $\tau$ with $j = 0.5$ (red, dashed curve), $j = 1$ (dot-dashed, magenta curve), and $j = 2$ (solid, blue curve). We have $s = 1$, $\varepsilon = 1$, $\omega_c = 10$, and $\Delta = 0.05$. The initial state is $\ket{j}$. }
\label{largejnocorrections}
\end{figure}

Let us now generalize the usual spin-boson model to deal with $N_S$ two-level systems interacting with a common environment. In this case, the system-environment Hamiltonian (in the `lab' frame) is given by \cite{VorrathPRL2005,ChaudhryPRA2013b,Chaudhryscirep2016}
$$ H_L = \varepsilon J_z + \Delta J_x + \sum_k \omega_k b_k^\dagger b_k + 2J_z \sum_k (g_k^* b_k + g_k b_k^\dagger), $$
where $J_{x,y,z}$ are the usual angular momentum operators obeying the commutation relations $[J_k, J_l] = i\varepsilon_{klm}J_m$. We now start from the spin coherent state $\ket{j}$ such that $J_z\ket{j} = j\ket{j}$ with $j = N_S/2$. Other eigenstates of $J_z$ can be considered as the initial state in a similar manner. Our objective is to again perform repeated projective measurements, described by the projector $\ket{j}\bra{j}$, with time interval $\tau$ and thereby investigate what happens to the effective decay rate. As before, the survival probability after one measurement is 
$$s(\tau) = \text{Tr}_{S,B} [\ket{j}\bra{j} e^{-iH_L \tau} \rho_L(0) e^{iH_L \tau}].$$
Since we consider the system and the environment to be strongly interacting, we once again perform the polaron tranformation given by $H = e^{\chi J_z} H_L e^{-\chi J_z}$, with $\chi$ the same as before. Then, we find that 
$$ H = \varepsilon J_z + \sum_k \omega_k b_k^\dagger b_k - \kappa J_z^2 + \frac{\Delta}{2}(J_+ X + J_- X^\dagger), $$
where $\kappa = 4\sum_k \frac{|g_k|^2}{\omega_k}$, and $J_\pm = J_x \pm iJ_y$ are the standard raising and lowering operators. Interestingly, the transformed Hamiltonian now contains a term proportional to $J_z^2$. This term arises because the collection of two-level systems interacting with the collective environment are indirectly interacting with each other. This term is obviously proportional to the identity operator for a single two-level system, and thus has no influence for a single two-level system. If the tunneling amplitude is small, then we again use perturbation theory and assume that, in the polaron frame, the system-environment correlations can be neglected. We find that now the effective decay rate is (see the Methods section)
\begin{equation}
\Gamma(\tau) = \frac{\Delta^2 j}{\tau} \int_0^\tau dt \int_0^t dt' e^{-\Phi_R(t')} \cos[\varepsilon t' + \kappa(1 - 2j)t' - \Phi_I(t')]. 
\label{decayratelargej}
\end{equation}
Here $\Phi_R(t)$ and $\Phi_I(t)$ are the same as defined before. This result obviously agrees with the result that we obtained for a single two-level system. Moreover, it is clear from Eq.~\eqref{decayratelargej} that increasing the system-environment coupling strength $G$ should reduce the effective decay rate due to the $e^{-\Phi_R(t')}$ factor in the integrand. This is precisely what we observe in Fig.~\ref{largejnocorrections}(a). Furthermore, it may be thought that increasing $j$ (or, equivalently, $N_S$) increases the effective decay rate. On the other hand, the dependence on $j$ is not so clear because of the presence of the indirect interaction. Namely, increasing $j$ increases the oscillatory behavior of the integrand due to the dependence of the integrand on $\cos[\varepsilon t' + \kappa(1 - 2j)t' - \Phi_I(t')]$. Thus, once the integral over this rapidly oscillating integrand is taken, we can again get a small number. Such a prediction is borne out by Fig.~\ref{largejnocorrections}(b) where the effective decay rate has been plotted for different values of $j$. It is obvious that there is a big difference between the single two-level system case and the more than one two-level system case. Furthermore, it seems that increasing $j$ can largely reduce the value of the effective decay rate, meaning that in the strong coupling regime, the indirect interaction helps in keeping the quantum state alive. 

\begin{figure}
\centering
{\includegraphics[scale = 0.5]{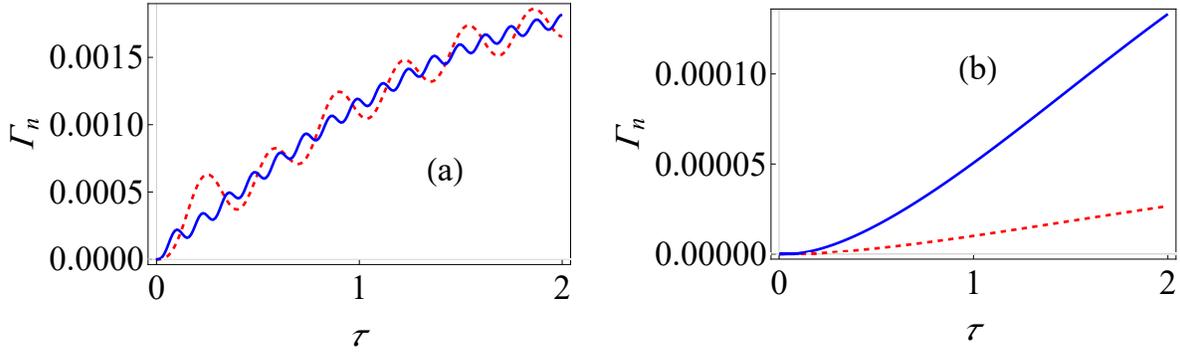}}
\caption{\textbf{Variation of the effective decay rate $\Gamma_n(\tau)$ for the large spin-boson model.} (a) Graph of $\Gamma$ (at zero temperature)  with modification for the strong-coupling regime as a function of $\tau$ with $j = 1$ for the system-environment coupling strength $G = 1$ (red, dashed curve) and $G = 2.5$ (solid, blue curve). Here we have used an Ohmic environment ($s = 1$), with $\varepsilon = 1$, $\omega_c = 10$, and $\Delta = 0.05$. The initial state is $\ket{j}$. (b) Behaviour of $\Gamma$ (at zero temperature) for the usual weak system-environment coupling scenario with $\varepsilon = 1$ and $\Delta = 0.05$ with $G = 0.001$ (dashed, red curve) and $G = 0.005$ (solid, blue curve). We have used again an Ohmic environment, the initial state is still $\ket{j}$ and $\omega_c = 10$.}
\label{largejwithcorrections}
\end{figure}

Let us now consider the situation where the evolution to the system Hamiltonian $H_{S,L} = \varepsilon J_z + \Delta J_x$ is removed before each measurement. In the polaron frame $H_{S,L}$ becomes $H_{S,P} = \varepsilon J_z + \frac{\Delta}{2}(J_+ X + J_- X^\dagger)$. The major difference now compared to the previous single two-level system case is that the total system-environment Hamiltonian in the polaron frame $H = H_{S,P} + H_B - \kappa J_z^2$ contains a term (namely, $-\kappa J_z^2$) that is not part of the system Hamiltonian in the polaron frame. As a result, when the system evolution is removed just before performing each measurement, the evolution induced by this extra term survives. Keeping this fact in mind, the effective decay rate $\Gamma_n(\tau)$ is now 
\begin{align}
\Gamma_n(\tau) =  \Gamma(\tau) + \Gamma_{\text{mod}}(\tau), 
\label{modifieddecayratelargej}
\end{align}
where $\Gamma(\tau)$ is given by Eq.~\eqref{decayratelargej} and 
\begin{align*}
\Gamma_{\text{mod}}(\tau) = \frac{\Delta^2}{\tau}(2j)\left\lbrace \frac{1}{\varepsilon^2} \sin^2 \left(\frac{\varepsilon \tau}{2}\right) e^{-\Phi_R(0) - i\Phi_I(0)} - \frac{1}{\varepsilon} \sin\left( \frac{\varepsilon \tau}{2}\right) \int_0^\tau dt e^{-\Phi_R(t)} \cos\left[\kappa(2j - 1)(2\tau - t) + \varepsilon\left(t - \frac{\tau}{2}\right) - \Phi_I(t)\right]\right\rbrace. 
\end{align*}
In Fig.~\ref{largejwithcorrections}(a), we have shown the behavior of $\Gamma_n(\tau)$ when the system-environment coupling strength is increased for $N_S = 2$. It should be obvious that we observe multiple Zeno-anti Zeno regimes. Also, increasing the coupling strength does not generally  increase the effective decay rate $\Gamma_n(\tau)$. This behavior should be contrasted with the weak coupling scenario. For weak coupling, it has been found that the effective decay rate is still given by Eq.~\eqref{weakcouplingdecay}, but now the filter function is $N_S$ times the filter function given by Eq.~\eqref{weakcouplingfilterfunction}\cite{Chaudhryscirep2016}. Thus, increasing the coupling strength should now increase the effective decay rate. This is precisely what is observed in Fig.\ref{largejwithcorrections}(b). Consequently, the weak coupling and the strong coupling regimes are very different for the strong and the weak coupling regimes. The difference is again due to the system-environment correlations.

\section*{Discussion} 

We have investigated the quantum Zeno and anti-Zeno effects for a single two-level system interacting strongly with an environment of harmonic oscillators. Although it seems that perturbation theory cannot be applied, we have applied a polaron transformation that can make the coupling strength effectively small in the transformed frame and thereby validate the use of perturbation theory. We have obtained general expressions for the effective decay rate, independent of any particular form of the spectral density of the environment. Thereafter, we have shown that the strong coupling regime shows both qualitative and quantitative differences in the behavior of the effective decay rate as a function of the measurement interval and the QZE to QAZE transitions as compared with the weak system-environment coupling scenario. The effective decay rate is no longer an overlap integral of the spectral density of the environment and some other function. Rather, there is a very pronounced non-linear dependence on the spectral density of the environment. Most importantly, increasing the coupling strength in the strong coupling regime can actually reduce the effective decay rate. These differences can be understood in terms of the significant role played by the system-environment correlations. Moreover, we have extended our results to many two-level systems interacting with a common environment. Once again, we obtained expressions for the effective decay rate that are independent of the spectral density of the environment. We illustrated that in this case as well the behavior of the effective decay rate is very different from the commonly considered weak coupling regime. Our results should be important for understanding the role of repeated measurements in quantum systems that are interacting strongly with their environment.

\section*{Methods}

\subsection*{The polaron transformation}

For completeness, let us sketch how to transform the spin-boson Hamiltonian to the polaron frame \cite{SilbeyJCP1984,VorrathPRL2005,ChinPRL2011,LeeJCP2012,LeePRE2012,GuzikJPCL2015}. We need to find $H = e^{\chi \sigma_z/2} H_L e^{-\chi \sigma_z/2}$. We use the identity 
$$ e^{\theta A} B e^{-\theta A} = B + \theta [A, B] + \frac{\theta^2}{2!}[A,[A,B]] + \hdots $$
Now, it is clear that $[\chi \sigma_z/2, \sigma_z/2] = 0$. Also, $\left[\sum_k \left(\frac{g_k}{\omega_k}b_k^\dagger - \frac{g_k^*}{\omega_k}b_k\right), \sum_k \omega_k b_k^\dagger b_k \right] = -\sigma_z \sum_k (g_k b_k^\dagger + g_k^* b_k).$ Carrying on, we find that $\left[\sigma_z \chi/2, \sigma_z \sum_k (g_k^* b_k + g_k b_k^\dagger)\right] = -2 \sum_k \frac{|g_k|^2}{\omega_k}. $ This is simply a c-number, so the higher-order commutators are zero. Furthermore, this c-number leads to a constant shift in the transformed Hamiltonian, and can thus be dropped. Putting all the commutators together, we find that 
$$e^{\chi \sigma_z/2} \left[ \frac{\varepsilon}{2} \sigma_z + \sum_k \omega_k b_k^\dagger b_k + \sigma_z \sum_k (g_k^* b_k + g_k b_k^\dagger) \right] e^{-\chi \sigma_z/2} = \frac{\varepsilon}{2}\sigma_z + \sum_k \omega_k b_k^\dagger b_k. $$
Next, we observe that $e^{\chi \sigma_z/2} \frac{\Delta}{2}\sigma_x e^{-\chi \sigma_z/2} = e^{\chi \sigma_z/2} \frac{\Delta}{2}(\sigma_+ + \sigma_-) e^{-\chi \sigma_z/2}$, where $\sigma_+$ and $\sigma_-$ are the standard spin half raising and lowering operators. Furthermore, $[\chi \sigma_z/2, \sigma_+] = \sigma_+ \chi$, leading to $e^{\chi \sigma_z/2}\sigma_+ e^{-\chi \sigma_z/2} = \sigma_+ e^\chi$. Similarly, $e^{\chi \sigma_z/2}\sigma_- e^{-\chi \sigma_z/2} = \sigma_- e^{-\chi}$. Thus, we finally have the required Hamiltonian in the polaron frame. 

For the large spin case, the calculation is very similar \cite{VorrathPRL2005}. The major difference is that now the c-number term that we dropped before cannot be dropped any longer since this term is proportional to $J_z^2$ (for the spin half case, this is proportional to the identity operator, so this is just a constant shift for the spin half case). Namely, we now find that 
$$ [\chi J_z, \varepsilon J_z + \sum_k \omega_k b_k^\dagger b_k + 2J_z \sum_k (g_k^* b_k + g_k b_k^\dagger)] = -2J_z\sum_k (g_k b_k^\dagger + g_k^* b_k) -8J_z^2 \sum_k \frac{|g_k|^2}{\omega_k}. $$
Also,
$$ [\chi J_z, -2J_z\sum_k (g_k b_k^\dagger + g_k^* b_k) -8J_z^2 \sum_k \frac{|g_k|^2}{\omega_k}] = 8J_z^2 \sum_k \frac{|g_k|^2}{\omega_k}. $$
The rest of the calculation is very similar to the spin half case, and leads to the Hamiltonian in the polaron frame.

\subsection*{Finding the system density matrix in the polaron frame}

Here we describe how to obtain the system density matrix in the polaron frame $\rho_S(\tau)$ just before performing the measurement at time $\tau$. We define $U_{\text{tot}}(\tau) = e^{-iH\tau} = U_0(\tau) U_I (\tau)$, where $U_0(\tau)$ is the unitary time-evolution operator corresponding to $H_S$ and $H_B$, while $U_I(\tau)$ is the `left over' part that we can find using time-dependent perturbation theory. Writing the system-environment coupling in the polaron frame as $\sum_\mu F_\mu \otimes B_\mu$, with $F_1 = \frac{\Delta}{2}\sigma_+$, $B_1 = X$, $F_2 = \frac{\Delta}{2}\sigma_-$ and $B_2 = X^\dagger$, $U_I(\tau)$ can be found to be $U_I(\tau) = \mathds{1} + A_1 + A_2$, with $A_1 = -i \sum_\mu \int_0^\tau \widetilde{F}_\mu(t_1) \widetilde{B}_\mu(t_1) dt_1$ and $A_2 = -\sum_{\mu \nu} \int_0^\tau dt_1 \int_0^{t_1} dt_2 \widetilde{F}_\mu(t_1) \widetilde{F}_\nu(t_2) \widetilde{B}_\mu(t_1) \widetilde{B}_\nu(t_2)$. Correct to second order in the tunneling amplitude $\Delta$, we can then write 
		\begin{align}
		\rho_S(\tau) \approx \, &\text{Tr}_B \lbrace U_0(\tau) [\rho_{\text{tot}}(0) + \rho_{\text{tot}}(0) A_1^\dagger + \rho_{\text{tot}}(0)A_2^\dagger + A_1 \rho_{\text{tot}}(0) + A_2 \rho_{\text{tot}}(0) + A_1 \rho_{\text{tot}}(0) A_1^\dagger] U_0^\dagger(\tau) \rbrace,
		\label{simplifythis}
		\end{align}
		where $\rho_{\text{tot}}(0) = \rho_S(0) \otimes \rho_B$. Eq.~\eqref{simplifythis} can now be simplified term by term. First, we find that $ \text{Tr}_B \lbrace U_0(\tau) \rho_{\text{tot}}(0)U_0^\dagger(\tau)\rbrace = \widetilde{\rho}_S(\tau)$,
			where $\widetilde{\rho}_S(\tau) = U_S(\tau)\rho_S(0)U_S^\dagger(\tau)$ is the system density matrix if the tunneling amplitude is zero. Next, we find that $\text{Tr}_B \lbrace U_0(\tau) \rho_{\text{tot}}(0)A_1^\dagger U_0^\dagger(\tau)\rbrace = i \sum_\mu \int_0^\tau dt_1 \, U_S(\tau) \rho_S(0) \widetilde{F}_\mu(t_1) U_S^\dagger(\tau) \langle \widetilde{B}_\mu (t_1) \rangle_B$, where $\langle \widetilde{B}_\mu (t_1) \rangle_B = \text{Tr}_B \lbrace U_B(\tau) \rho_B \widetilde{B}_\mu(t_1) U_B^\dagger(\tau)\rbrace$. Similarly, $\text{Tr}_B \lbrace U_0(\tau) A_1 \rho_{\text{tot}}(0) U_0^\dagger(\tau)\rbrace = -i \sum_\mu \int_0^\tau dt_1 \, U_S(\tau) \widetilde{F}_\mu(t_1)\rho_S(0)  U_S^\dagger(\tau) \langle \widetilde{B}_\mu (t_1) \rangle_B$. Carrying on, 
			\begin{equation*} 
			\tr_B \lbrace U_0(\tau) A_2 \tot{\rho}(0) U_0^\dagger (\tau) \rbrace = 
 -\sum_{\mu \nu} \int_0^\tau dt_1 \int_0^{t_1} dt_2 \, U_S(\tau) \widetilde{F}_\mu(t_1) \widetilde{F}_\nu(t_2) \rho_S(0) U_S^\dagger(\tau) C_{\mu \nu}(t_1,t_2),
			\end{equation*}
			with the environment correlation function $C_{\mu \nu}(t_1,t_2)$ defined as $C_{\mu \nu}(t_1,t_2) = \langle \widetilde{B}_\mu(t_1)\widetilde{B}_\nu(t_2)\rangle_B = \tr_B \lbrace \widetilde{B}_\mu(t_1)\widetilde{B}_\nu(t_2)\rho_B\rbrace$. Similarly, 
			\begin{equation*}
			\tr_B \lbrace U_0(\tau) \tot{\rho}(0) A_2^\dagger U_0^\dagger (\tau) \rbrace = -\sum_{\mu \nu} \int_0^\tau dt_1 \int_0^{t_1} dt_2 \, U_S(\tau) \rho_S(0) \widetilde{F}_\nu(t_2) \widetilde{F}_\mu(t_1) U_S^\dagger(\tau) C_{\nu \mu}(t_2,t_1).
			\end{equation*}	
			Finally, 
			\begin{equation*}
			\tr_B\lbrace U_0(\tau)A_1 \tot{\rho}A_1^\dagger U_0^\dagger(\tau)\rbrace = 
			\sum_{\mu \nu} \int_0^\tau dt_1 \int_0^\tau dt_2 U_S(\tau) \widetilde{F}_\mu(t_1)\rho_S(0)\widetilde{F}_\nu(t_2)U_S^\dagger(\tau) C_{\nu \mu}(t_2,t_1).
			\end{equation*}
			Using the fact that $\int_0^\tau dt_1 \int_0^\tau dt_2 = \int_0^\tau dt_1 \int_0^{t_1} dt_2 + \int_0^\tau dt_2 \int_0^{t_2} dt_1$, 
			\begin{align*}
			\tr_B\lbrace U_0(\tau)A_1 \tot{\rho}A_1^\dagger U_0^\dagger(\tau)\rbrace = 
			\sum_{\mu \nu} \int_0^\tau dt_1 \int_0^{t_1} dt_2 U_S(\tau) \widetilde{F}_\mu(t_1)\rho_S(0)\widetilde{F}_\nu(t_2)U_S^\dagger(\tau) 
			C_{\nu \mu} (t_2, t_1) + \text{h.c.},
			\end{align*}
			where h.c. denotes hermitian conjugate. Putting all the terms back together, the system density matrix can be written as Eq.~\eqref{densitymatrixattau}. 
			
		\subsection*{Finding the effective decay rate}
		We now explain how to find the effective decay rate given by Eq.~\eqref{decayrateoneTLS}. 
			With the system density matrix at time $\tau$ available, we first calculate the survival probability $s(\tau)$. This can be done via 
			$s(\tau) = 1 - \opav{\downarrow}{\rho_S(\tau)}{\downarrow}$. Since the state $\ket{\downarrow}$ is an eigenstate of $H_S$, and $\rho_S(0) = \ket{\uparrow}\bra{\uparrow}$, it is straightforward to see that 
			\begin{align*}
			s(\tau) &= 1 - 2 \text{Re} \biggl( \sum_{\mu \nu}\int_0^\tau dt_1 \int_0^{t_1} dt_2  C_{\mu \nu}(t_1, t_2) 
			\opav{\downarrow}{[\widetilde{F}_\nu(t_2)\rho_S(0),\widetilde{F}_\mu(t_1)]}{\downarrow} \biggr) \notag \\
			&= 1 - 2 \text{Re} \biggl(\sum_{\mu \nu} \int_0^\tau dt_1 \int_0^{t_1} dt_2  C_{\mu \nu}(t_1, t_2) 
			\opav{\downarrow}{\widetilde{F}_\nu(t_2)}{\uparrow}\opav{\uparrow}{\widetilde{F}_\mu(t_1)}{\downarrow} \biggr).
			\end{align*}
			We now note that, since $F_1 = \frac{\Delta}{2}\sigma_+$, $F_2 = \frac{\Delta}{2}\sigma_-$, and $H_S = \frac{\varepsilon}{2}\sigma_z$, $\widetilde{F}_1(t) = \frac{\Delta}{2} \sigma_+ e^{i\varepsilon t}$ and $\widetilde{F}_2(t) = \frac{\Delta}{2} \sigma_- e^{-i\varepsilon t}$. Therefore,
			\begin{align*}
			s(\tau) = 1 - 2 \text{Re} \biggl(\int_0^\tau dt_1 \int_0^{t_1} dt_2  C_{12}(t_1, t_2) 
			\opav{\downarrow}{\widetilde{F}_2(t_2)}{\uparrow}\opav{\uparrow}{\widetilde{F}_1(t_1)}{\downarrow} \biggr) = 1 - \frac{\Delta^2}{2} \text{Re} \biggl(\int_0^\tau dt_1 \int_0^{t_1} dt_2  C_{12}(t_1, t_2) 
			e^{i\varepsilon(t_1 - t_2)} \biggr).		
			\end{align*}
			What remains to be worked out is the environment correlation function $C_{12}(t_1,t_2) = \tr_B [\rho_B \widetilde{X}(t_1) \widetilde{X}^\dagger(t_2)]$. Using the cyclic invariance of the trace, it is clear that this correlation function is actually only a function of $t_1 - t_2$ only, since $\tr_B [\rho_B \widetilde{X}(t_1) \widetilde{X}^\dagger(t_2)] = \tr_B [\rho_B e^{iH_B(t_1 - t_2)}Xe^{-iH_B(t_1 - t_2)}X^\dagger] = C_{12}(t_1 - t_2)$. Thus, 
		\begin{align*}
			s(\tau) = 1 - \frac{\Delta^2}{2} \text{Re} \biggl(\int_0^\tau dt_1 \int_0^{t_1} dt_2  C_{12}(t_1 - t_2) 
			e^{i\varepsilon(t_1 - t_2)} \biggr) = 1 - \frac{\Delta^2}{2} \text{Re} \biggl(\int_0^\tau dt \int_0^{t} dt'  C_{12}(t') 
			e^{i\varepsilon t'} \biggr)	,	
			\end{align*}	
			where we have introduced $t' = t_1 - t_2$. 
			The calculation of $C_{12}(t')$ can be performed as follows. First, we use the useful fact that for $\tr_B[\rho_B e^Z] = e^{\langle Z^2 \rangle/2}$ where $Z$ is a linear function of the creation and annihilation operators. Second, to obtain a single exponential so that the previous identity can be used, we use the identity that for any two operators $X$ and $Y$, $e^X e^Y = e^{X+Y+\frac{1}{2}[X,Y] \hdots}$. Fortunately for us, the series terminates for our case, so the higher order terms are zero. Using these two identities, we find that $C_{12}(t') = e^{-\Phi_R(t')} e^{-i\Phi_I(t')}$ where $\Phi_R(t)$ and $\Phi_I(t)$ have been defined in Eq.~\eqref{PhiRandPhiI}, and the spectral density of the environment has been introduced as $\sum_k |g_k|^2 (\hdots) \rightarrow \int_0^\infty d\omega J(\omega)(\hdots)$. This finally leads to 
			$s(\tau) = 1 - \frac{\Delta^2}{2} \int_0^\tau dt \int_0^t dt'  e^{-\Phi_R(t')} \cos[\varepsilon t' - \Phi_I(t')]$. We can define an effective decay rate $\Gamma(\tau) = -\frac{1}{\tau} \ln s(\tau)$. For small $\Delta$, we expect the deviation of the survival probability from one to be small. Thus, we end up with Eq.~\eqref{decayrateoneTLS}.

\subsection*{Calculating the modified decay rate}


Let us now briefly sketch how to obtain Eq.~\eqref{modifieddecayratesingleTLS}. We note that the system Hamiltonian $H_{S,L}$ becomes in the polaron frame $H_{S,P} = \frac{\varepsilon}{2}\sigma_z + \frac{\Delta}{2}(\sigma_+ X + \sigma_- X^\dagger)$. Then, to second order in $\Delta$, $ e^{-iH_{S,P}\tau} \approx \mathds{1} + A_{SP}^{(1)} + A_{SP}^{(2)}$, with $  A_{SP}^{(1)} = -i \int_0^\tau dt_1 \sum_{\mu} \widetilde{F}_\mu (t_1) B_\mu$, and $  A_{SP}^{(2)} = -\int_0^\tau dt_1 \int_0^\tau dt_2 \sum_{\mu \nu} \widetilde{F}_\mu (t_1) B_\mu  \widetilde{F}_\nu (t_2) B_\nu$. Here, $ \widetilde{F}_\mu (t) = e^{i\varepsilon \sigma_z t/2}  F_\mu  e^{-i\varepsilon \sigma_z t/2}$. Substituting these expressions in the expression for the survival probability as well as the perturbation expansions for $e^{iH\tau}$ and $e^{-iH\tau}$, and keeping terms up to second order in $\Delta$, we find that the new survival probability consists of the previous survival probability plus some additional terms. It can be be easily seen that most of these additional terms, once the trace with the projector $\ket{\downarrow}\bra{\downarrow}$ is taken, give zero. The additional terms that need to be worked out are $\tr_B [\langle \downarrow|U_B(\tau) A_1 \rho_S(0) \rho_B U_B^\dagger (\tau) A_{SP}^{(1)}|\downarrow \rangle]$, $\tr_B [\langle \downarrow| A_{SP}^{(1)\dagger} U_B(\tau) \rho_S(0) \rho_B A_1^\dagger  U_B^\dagger (\tau)|\downarrow\rangle]$, and $\tr_B [\langle \downarrow| A_{SP}^{(1)\dagger} U_B(\tau) \rho_S(0) \rho_B U_B^\dagger A_{SP}^{(1)}|\downarrow\rangle]$. The first of these terms is equal to 
$$ - \sum_{\mu \nu} \int_0^\tau dt_1 \int_0^\tau dt_2 \opav{\downarrow}{\widetilde{F}_\nu(t_1) \rho_S(0) \widetilde{F}_\mu(t_2)}{\downarrow} C_{\mu \nu} (\tau - t_1), $$
while the second is simply the hermitian conjugate of the first. On the other hand, the last term is equal to 
$$ \sum_{\mu \nu} \int_0^\tau dt_1 \int_0^\tau dt_2 \opav{\downarrow}{\widetilde{F}_\nu(t_1) \rho_S(0) \widetilde{F}_\mu(t_2)}{\downarrow}C_{\mu \nu}(0). $$ 
Next, we use the fact that $F_1 = \frac{\Delta}{2}\sigma_+$ and $F_2 = \frac{\Delta}{2}\sigma_-$ to simply the inner products. Putting all the pieces together, we arrive at Eq.~\eqref{modifieddecayratesingleTLS}. The calculation of $\Gamma_n(\tau)$ for the large spin case is quite similar. One only needs to be careful about the fact that the system-environment Hamiltonian, in the polaron frame, contains a term, namely $-\kappa J_z^2$, that is not a part of the transformed system Hamiltonian $H_{S,P}$.

\bibliography{thesisbib}

\end{document}